\documentclass[aip,reprint,twocolumn,amsmath,amssymb,groupedaddress]{revtex4-1}

\usepackage{graphicx,hyperref,units}
\usepackage{amsmath,amssymb,lmodern,mathrsfs}

\newcommand{\bra}[1]{\ensuremath{\left\langle{#1}\right\vert}}

\newcommand{\ket}[1]{\ensuremath{\left|{#1}\right\rangle}}

\newcommand{\pvec}[1]{\vec{#1}\mkern2mu\vphantom{#1}}
\begin{document}

\title{Optical Signatures of Quantum Delocalization over Extended Domains in Photosynthetic Membranes}
\author{Christopher A. Schroeder$^{1,2,*}$, Felipe Caycedo-Soler$^{1,*}$, Susana F. Huelga $^1$, and Martin B. Plenio$^1$}
\affiliation{$^*$ These authors contributed equally.}
\affiliation{$^1$ Institute of Theoretical Physics, University of Ulm, Albert-Einstein-Allee 11 D - 89069 Ulm, Germany}
\affiliation{$^2$ Joint Quantum Institute, Department of Physics, University of Maryland and National Institute of Standards and Technology, College Park, MD 20742, USA}

\begin{abstract}
The prospect of coherent dynamics and excitonic delocalization across several light-harvesting structures in photosynthetic membranes is of considerable interest, but challenging to explore experimentally. Here we demonstrate theoretically that the excitonic delocalization across extended domains involving several light-harvesting complexes can lead to unambiguous signatures in the optical response, specifically, linear absorption spectra. We characterize, under experimentally established conditions of molecular assembly and protein-induced inhomogeneities, the optical absorption in these arrays from polarized and unpolarized excitation, and demonstrate that it can be used as a diagnostic tool to determine the resonance coupling between iso-energetic light-harvesting structures. The knowledge of these couplings would then provide further insight into the dynamical properties of transfer, such as facilitating the accurate determination of F\"orster rates.
\end{abstract}

\maketitle






\section{Introduction}
Nature has evolved a variety of photosynthetic architectures. A detailed, quantitative understanding of the principles that underly their function could assist the design of future energy conversion devices. A wealth of careful structural and spectral studies \cite{Vischers_Biochem_1991, karrasch95, Bahatyrova_2004Nature, scheuring2006, shreve1991, Bullough2004} complemented by first principle calculations \cite{Hu_2002, Strumpfer_JCP2012, silbey_PRL_2006, Strumpfer_JPC2012} have contributed to our current understanding of exciton transfer dynamics in light-harvesting (LH) antenna and reaction center (RC) pigment-protein complexes. Recently, interest in this topic has intensified due to observations of persistent oscillatory features in non-linear optical experiments\cite{Romero14, Fuller14, Flemming_2007Science, Engel2010, Engel_Nature2006, Flemming2005, EngelNJP}, which have been reproduced in various LH structures. This has motivated work which reevaluates the nature of the interaction between excitonic dynamics and vibrational motion\cite{Mohseni2008, Pleniol_NJP_2008, Olaya_2008}.

\begin{figure*}[htbp!]
\includegraphics[width=16cm]{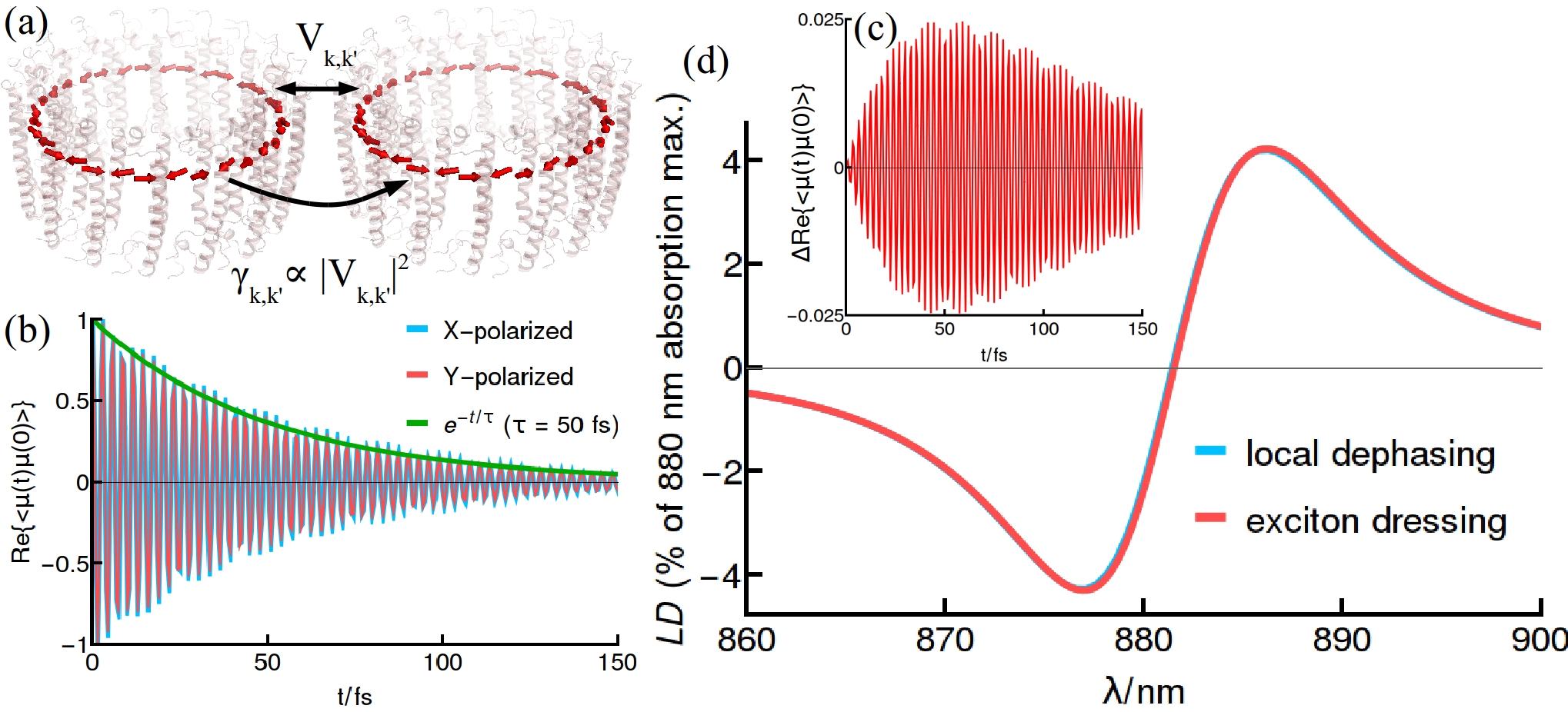}
\caption{(a) The $Q_y$ transition dipole moments (red arrows) of the BChl pigments and the protein scaffold (light red) of the LH1 complex in {\it R. rubrum}, are pictured schematically\cite{Autenrieth_2002}. The rate of incoherent energy transfer between excitons $k$ and $k'$ on different LH complexes, $\gamma_{k,k'}$, is determined by the resonance coupling $V_{k,k'}$, as explained in the main text. (b) The oscillating dipole-dipole correlation function (DDCF) contains all signatures of the resonance coupling and decays due to decoherence, which leads to homogeneous spectral broadening. The difference between the $\hat{x}$- and $\hat{y}$-polarized correlation functions (c) contains the signatures of coupling between rings, $V_{k,k'}$, and can be obtained experimentally by measuring (d) the linear dichroism ($LD$), which is the difference between polarized absorption along these orthogonal axes. Importantly, the effects of local dephasing on the $LD$ can be reproduced by both the DDCF, Eq. 2, or a simple dressing of excitonic states according to Eq. \ref{Aspec2}.}
\label{fig1}
\end{figure*}

Recent work has examined excitonic delocalization restricted to closely-packed pigments in single LH units \cite{Renger_2006_BioJ, Strumpfer_JPC2009, Strumpfer_JCP2012}. This approximation is valid for studies of inter-complex energy transfer and fluorescence, which take place on time scales ($\sim$ ps--ns) much longer than the dephasing ($\sim$ 100 fs), which reduces the excitons to localized (single-unit) domains \cite{Strumpfer_JPC2012, Timpmann_ChemPhysLett2005, Trinkunas_JLum_2006}. 
However, as we will show, for processes that are characterized through observables with a faster built-in time-scale, such as absorption, which depends on the dipole-dipole correlation function (DDCF) oscillating at optical frequencies, extended delocalization across LH complexes must be taken into account in an accurate description \cite{Caycedo14}.
Here, we propose and characterize theoretically an experimental scheme, based on simple linear absorption measurements, which quantifies the resonance coupling among LH complexes in purple bacteria. Based on firm theoretical analysis, we show how this information facilitates the determination of incoherent excitonic transfer rates in systems where the donor's fluorescence and acceptor's absorption overlap.

\section{Materials and Methods}
{\bf Purple bacteria.} Photosynthetic membranes of purple bacteria are composed, in general, of 
 two-dimensional arrays of LH1 and LH2 complexes, which are responsible for the absorption of light and the subsequent transport of excitation energy to RC pigments, where charge separation drives metabolism under photosynthetic growth conditions\cite{Hu_2002,Bahatyrova_2004Nature, scheuring2005b, scheuring2006, Scheuring_2004_PNAS, Bullough2004, karrasch95, Walz97, Caycedo_2010PRL}. In such membranes, excitonic energy transfer among LH1 and LH2 units has been characterised in terms of incoherent (thermal) hopping \cite{Ritz01}. However, for a description of absorption, which 
has a much faster built-in time-scale 
than inter-complex energy transfer and dephasing, excitonic delocalization across several harvesting complexes may become relevant. In this work, we analyse excitonic delocalisation across many LH complexes. We will show that this delocalization leads to a redistribution of absorption intensity which is experimentally accessible by exploiting the symmetry of LH1 and LH2 complexes.

For concreteness, we develop our analysis using the LH1 of {\it R. rubrum}, whose assumed homology of LH1 to LH2 harvesting units, permits a straightforward generalization to membranes composed of either complex. The available X-ray structure of the LH1 complex from {\it Tch. tepidum}\cite{Miki_2014Nature} and vanishing fluorescence anisotropy of LH1s in {\it R. rubrum} \cite{Ghosh_Biochem_2003} provide support for a closed ring structure with dimeric repeating units of $2N=32$ bacteriochlorophyll (BChl) pigments in a $C_{16}$-fold symmetry, as depicted schematically in Figure \ref{fig1}(a), although not all LH1 complexes are closed or circular \cite{Cogdell_06}. LH1 complexes from {\it R. rubrum} naturally aggregate via protein domain-mediated interactions into arrays with a tetragonal packing and can be grown without the constituent RCs\cite{karrasch95}. Due to symmetry, the complex develops a single bright band at 880 nm, which, in the case of \emph{R. rubrum}, has 2.4 times the dipole strength from what is expected from the addition of 32 dipoles $\vec{d}$ \cite{Ghosh, Ghosh_88}. Such hyperchromism is accounted for by increasing the magnitude of the induced dipoles $|\vec{d}_i| \rightarrow |\sqrt{2.4} \times  \vec{d}|=\sqrt{2.4}  \times 6.3$~Debye $=9.8$~Debye. In {\it R. sphaeroides}, LH1 complexes tend to dimerize \cite{karrasch95,scheuring2006,Bahatyrova_2004Nature}. 

{\bf Exciton formalism.} Excitonic properties of membranes subject to low-intensity illumination can be obtained from the electronic Hamiltonian in the single excitation subspace:
\begin{align}
\mathscr{H} = \sum_m \epsilon_m \ket{m}\bra{m} + \sum_{m,n} J_{mn} \ket{n}\bra{m} \label{eq1}
\end{align}
where $\epsilon_m$ is the excitation energy of pigment $m$ and $J_{nm}$ is the coupling, through the Coulomb exchange mechanism, between the $Q_y$ induced transition dipoles, $\vec{d}_i$, of the electronic excited states $\ket{n}$ and $\ket{m}$ on pigments $n$ and $m$ \cite{PhotoEx}. Details of the coherent interaction lead to specific eigen-frequencies $\omega_{\alpha}$ from eigenstates $\ket{\alpha}=\sum_n c_n^\alpha\ket{n}$, in general, delocalized over the complete set of pigments accounted for (see Supplementary Information for Hamiltonian
parameters).  In general, the resonance coupling between different rings, labeled $V_{k,k'}$ in Fig.\ref{fig1}(a), is the origin of EET, which is well-described as an incoherent process between {\it different} rings (occurring at a rate $\gamma_{k,k'} \propto |V_{k,k'}|^2$)\cite{forster1965}.  The key prediction of this article, illustrated in Fig.\ref{fig1}, is that this resonance coupling introduces experimentally-accessible signatures in the linear absorption which evidence excitonic delocalization across different rings and can be used to obtain estimates of the EET rates $\gamma_{k,k'}$. 

The absorption spectrum $A_{\hat E}(\omega)$ along an excitation polarization direction $\hat E$ is obtained from the Laplace-Fourier transform of the DDCF, $\langle (\vec{D}(t)\cdot\hat E)(\vec{D}(0)\cdot\hat E)\rangle$ \cite{Tannor},
\begin{eqnarray}
A_{\hat{E}}(\omega)&=&\int_0^\infty \langle (\vec{D}(t)\cdot \hat E)(\vec{D}(0)\cdot \hat E)\rangle e^{i\omega t} dt\approx 
\int_0^\infty \mbox{Tr} \{ \left(e^{{\cal L} t}\vec{D}(0)\cdot \hat E\right)\vec{D}(0)\cdot \hat E\}e^{i\omega t} dt\label{Aspec}\\
&\approx& \sum_\alpha |\vec D_\alpha\cdot \hat E|^2 f(\omega-\omega_\alpha)\label{Aspec2}
\end{eqnarray}
The propagator  $e^{{\cal L} t}$ contains all excitonic couplings and dispersive processes of relevance for excitonic dynamics. Within the DDCF description, the evolution of the  dipole moment operator $\vec{ D}(t)=e^{{\cal L}\,t}\vec D(0)=e^{{\cal L}\,t}\sum_i \vec d_i\ket{m}\bra{0}$ leads to oscillations at optical frequencies $\propto e^{i\omega_\alpha t}$ with a period of about $\approx$ 2-3 femtoseconds, due to the Hamiltonian part of the propagator. These coherent signatures are degraded by environmental decoherence, which in the worst-case scenario (in terms of processes that degrade the excitonic delocalisation) may lead to independent fluctuations of pigment energies $\omega_i$ within the excitation's lifetime, local pure dephasing, occurring on a time-scale of $\gamma_d^{-1}\approx 50-100$ fs \cite{Engel_PNAS2012,Hildner2013}. 

The DDCF evolves for many tens of optical cycles, as seen in Fig. 1(b), before dephasing sets in, which can be sufficient to imprint in the absorption spectra features of excitonic coupling, i.e., excitonic delocalization among several LH units. Such features are difficult or impossible to observe in processes with a longer built-in timescale, such as EET. Our interest is to connect the optical signatures from this  DDCF function with excitonic delocalization across {\it multiple} rings. In order to be conservative, we initially consider the worst-case scenario for degradation of such extended excitonic delocalization, introduced by a local dephasing superoperator of the form ${\cal L}_d\rho = \sum_m\frac{\gamma_d}{2} (\sigma_z^m \rho \sigma_z^m - \rho)$, with $\sigma_z^m=\ket{m}\bra{m}-\ket{0}\bra{0}$ and the density operator $\rho$, consistent with random temporal fluctuations of the energy gap between levels $\ket{m}$ and $\ket{0}$. When derived from a microscopic approach, the considered dephasing model describes the action of a fully Markovian environment in the infinite temperature limit. The vectorial nature of the dipole moment $\vec{D}$ permits discrimination of absorption along the axis that joins two coherently-coupled rings and perpendicular to it, based upon the calculation of the respective DDCF of Fig. 1(b). The subtraction of these two signals, presented in Fig.1(c), shows that their DDCF differ, a fact made even more conspicuous by the subtraction of absorption spectra (through the Laplace-Fourier transform from Eq. \ref{Aspec}) corresponding to these orthogonal directions, namely the linear dichroism ($LD$). This result should be contrasted with the vanishing $LD$ of single circular rings, or equivalently, assemblies of rings where the resonance coupling between rings plays no role, as discussed below. As these calcuations show, strong local dephasing with a rate of $\gamma_d$=1/50 fs is not enough to smear out the finite $LD$ for the coupled rings, and displays the robustness of optical measures for understanding the extent of excitonic delocalization in photosynthetic membranes.  Reassuringly, the result obtained from the local dephasing model is virtually indistinguishable from the usual procedure illustrated by Eq. \ref{Aspec2}, in which excitonic energies $\omega_\alpha$ are used as the center of a dressing function $f(\omega-\omega_\alpha)$, whose weight in the complete absorption spectrum is determined by the exciton dipole strength $|\vec{D}_\alpha \cdot \hat{E}|^2$. This agreement justifies the reduction of the complexity of the DDCF calculation to a simple diagonalization of the full Hamiltonian and the use of Lorentzian line-shape functions $f(\omega)= \gamma_d^2 / (\omega^2+\gamma_d^2)$, in order to establish the influence of the electronic coherent coupling in polarized absorption spectra.  Moreover,  it emphasizes the necessity of the excitonic dipole strength $|\vec{D}_\alpha|^2$, and therefore the relevance/necessity of excitonic states $\ket{\alpha}$, for the description/understanding of the the absorption process in ring assemblies.


\section{Results and Discussion}

Excitons $\ket{\alpha}$ are affected by their interactions with the protein environment, which  dynamically degrade the electronic coherence  within the excitation lifetime (homogeneous broadening) captured by either ${\cal L}_{d}$ or the dressing procedure, and quasi-static fluctuations (inhomogeneous broadening) of the excitation energy and the Coulomb coupling.  An interplay between both types of broadening leads to the observed width of the absorption spectrum, which in the case of LH1 is approximately $\Gamma \approx$ 465 cm$^{-1}$.  If the homogeneous line shape function is consistent with a dephasing lifetime of $\gamma_d=1/50$ fs,  a standard deviation of a Gaussian pigment site-energy variation of 325 cm$^{-1}$ is required to recover the full width of the LH1 absorption (further details in the Supplementary Information). This latter value is in agreement with previous estimations  \cite{Monshouwer97, Timpmann_ChemPhysLett2005,  Scholes03, Freiberg92, Kramer84, Rijgersberg80}.

\begin{figure*}[t!]
\includegraphics[width=16cm]{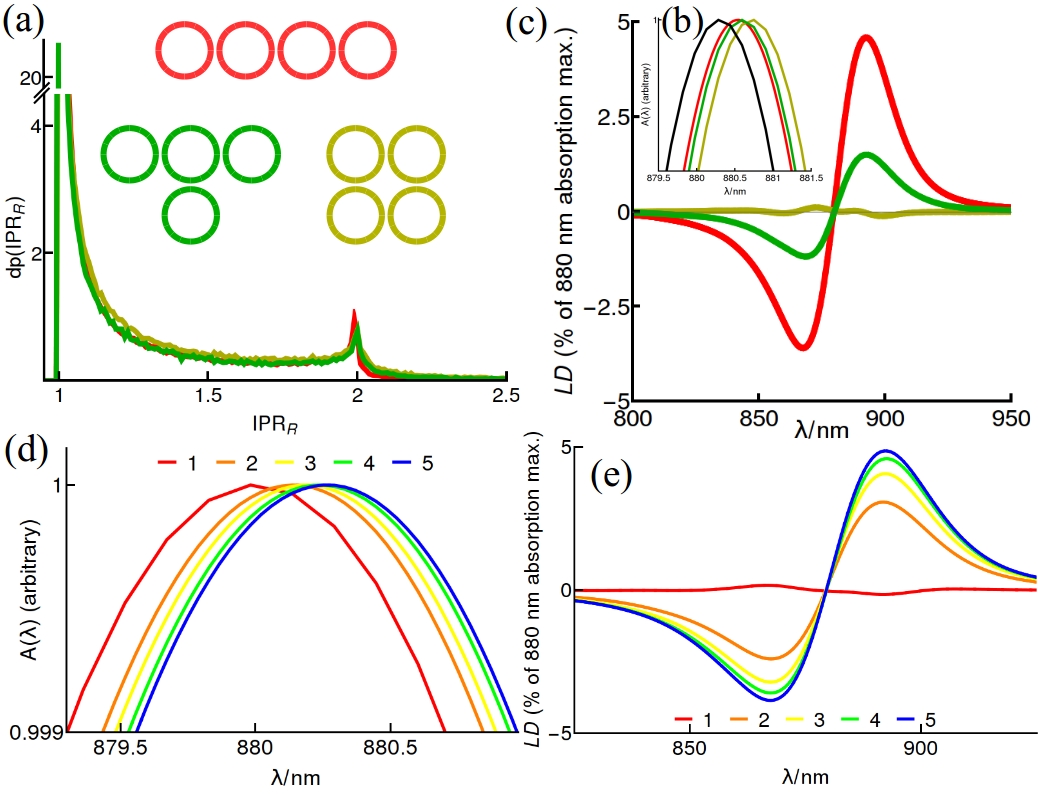}
\caption{Extended delocalization across multi-ring arrays in  different multiple-ring configurations. Each four-ring configuration considered in (a) exhibits a similar, although not identical, extent of delocalization, measured by the ring inverse participation ratio $IPR_R$. Plotted is the probability distribution, $dp(IPR_R)$, of the $IPR_R$ of the exciton state with greatest dipole strength, averaged over realisations of static disorder. The average delocalization length, $\langle IPR_R\rangle$, increases slightly with the number of neighbors, i.e. $\langle IPR_R\rangle=\{1.24, 1.27, 1.33\}$ for the line, T-, and square configurations, respectively. However, the delocalization is witnessed by linear dichroism (LD) only when the assembly exhibits an asymmetry, like in the T- and line assemblies, as shown in (c). (d) Red-shifted absorption and (e) non-zero linear-dichroism ({\it LD}) for linear arrays ($Q$=1, calculated for different values of $R=1,...,5$) witness extended delocalization over multiple LH1 complexes. Results from $5 \times 10^4$ realisations of site-energy variations with standard deviation 325 cm$^{-1}$ and $\gamma_d$ = 1/50 fs$^{-1}$.}
\label{fig1B}
\end{figure*}

{\bf Extended delocalization: Absorption and LD as a witness of delocalization in ring assemblies.} 
 The relevance of the excitonic dipole strength  $|\vec{D}_\alpha|^2$ for the description of the optical response of ring assemblies motivates our study of the excitons $\ket{\alpha}$ in multiple ring arrays. 

In order to estimate the influence of inhomogeneities on the extent of excitons, the usual inverse participation ratio \cite{Thouless_74,silbey_PRL_2006,Dahlbom01}, is generalised to quantify delocalisation {\it across} rings $R$ of a given exciton $\alpha$, namely the ring participation ratio $IPR_R^\alpha=1/\sum_{M=1}^R(\sum^{2N}_{n\epsilon M} |c_n^\alpha|^2)^2$. The $IPR_R$ ranges from 1 to $R$, and $IPR_R>1$ unequivocally represents excitonic delocalisation over more than a single ring. An estimation of the size of excitons in each of the three configurations presented in Fig.\ref{fig1B}(a) shows that absoprtion in these arrays generates excitons that in general extend over domains larger than a single ring. This figure also shows that excitons  have similar sizes for a fixed set of coherently coupled rings. In more detail, small differences emerge from the specific connectivity of each configuration, setting the square, T-like and linear configurations with exciton sizes in descending order. Importantly, the histogram of $IPR_R$ of Fig.\ref{fig1B} is made upon the two most optically active states from every noise realisation, and therefore represents the accesible states through optical excitation.

For uncoupled LH1 complexes, the circular symmetry leads to an optical response predominantly determined by two orthogonally-polarized degenerate exciton states and vanishing {\it LD} \cite{Caycedo14,Davydov,Ghosh_Biochem_2003,Wu97}. However, the coherent excitonic interaction bettween rings may lift such a symmetry and result in shifts in the unpolarised absoprtion spectrum and/or to a finite $LD$.  The key theoretical predictions of this article are summarized in Figures \ref{fig1B}(b)-(e). Namely, a shift in the unpolarised absorption spectra will generally be observed in assemblies of many rings, while  a non-zero {\it LD} will be measured in arrays with a large aspect ratio and small width, i.e. linear-like arrays.

A close inspection of the absorption spectrum (Fig.\ref{fig1B}(b)) of the configurations from the inset of Fig.\ref{fig1B}(a), shows very similar spectra for the T-like and linear configurations, while the square assembly does present a larger shift to the red than the former geometries, in comparison to the spectrum from a single ring. In practice, shifts that account for fractions of nanometers can be difficult to discriminate and a complementary measure, namely the $LD$, can be of help to understand whether the shifts observed are compatible with excitons formed across several rings. Even though all these structures present a finite shift with respect to the single ring spectrum, notice however, that not all multiple-ring arrangements present a finite $LD$. For instance the symmetric square assembly-- which presents the greatest $IPR_R$-- has vanishing $LD$, while the linear array shows a $LD$ with a contrast of $\simeq 4-5 \%$ of the absorption maximum. From these results it is apparent that, although each assembly presents similar delocalization and shifts in the unpolarised absorption, asymmetric assemblies provide a suitable scenario for the experimental observation of excitonic delocalization using polarized spectroscopy. 
When the focus is placed  in the linear-like arrays, it is observed that  unpolarised absorption spectra shows  greater shifts in  Fig.\ref{fig1B}(d) with increase of the number of rings in the linear array. Interestingly, the amplitude of the $LD$ also increases with the linear array size, from $\approx 2.5\%$ for 2 rings, rising up until it seems to saturate to $\approx 5\%$ for 5 rings.

It must be highlighted that just due to homogeneous broadening, the LD amplitude for 2 rings is about 5\% while if the inhomogeneities are included, it reduces to about 2.5\% of the total spectra. We have tested that this latter value is robust to  different compositions of homogeneous and inhomogeneous broadening restricted to the same full broadening of the optical transition. The inhomogeneities in a sample may not be restricted to environmentally-induced inhomogeneities in the pigment energies and can indeed be caused by structural perturbations, such as variations in the inter-complex distance or geometrical perturbations that produce ellipticity. A thorough study of both effects is presented in the Supplementary Information, from  where it can be concluded that the specific details of the broadening mechanisms do not play an important role for the determination of the $LD$ amplitude.

These full numerical simulations of the polarised and unpolarised absorption spectrum highlight 1) in general, (small) shifts in the absorption spectra are expected in arrays of rings; 2) a finite $LD$, in particular for linear-like assemblies, will be observed due to the coherent coupling of the rings, and 3) the contrast of the $LD$ depends inversely on the total broadening of the optical transition and not on the specific details of the broadening mechanism.

\begin{figure*}[t!]
\includegraphics[width=8cm]{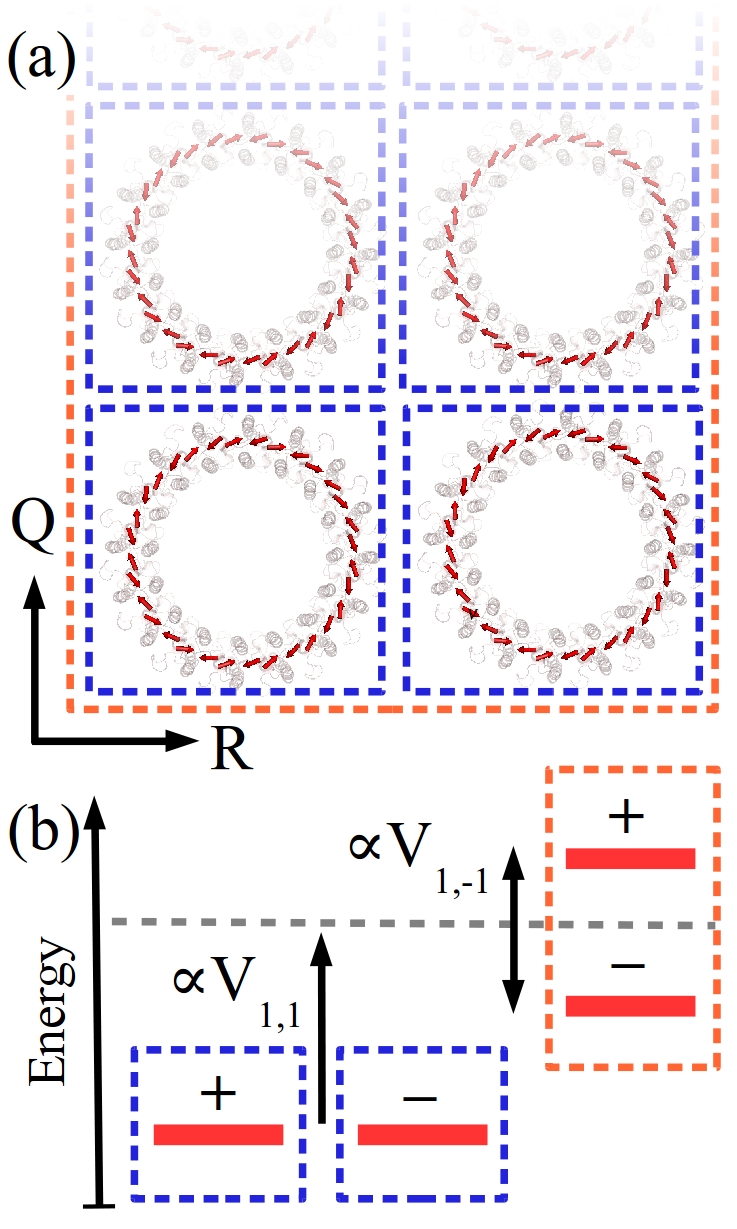}
\caption{(a) LH complexes in purple bacteria naturally arrange themselves in well-ordered 2D arrays. The spectral consequences of extended delocalization in such an arrangement can be understood by examining a rectangular $Q \times R$ array. The absorption anisotropy is strongest for linear-like arrays ($Q/R \gg 1$), represented schematically by the color gradient. (b) The circular symmetry of single LH complexes concentrates the optical absorption strength in degenerate states with orthogonal polarization (blue bounding boxes). The circular symmetry of the combined system is broken upon interaction (orange bounding boxes), leading to a measurable splitting of polarized states.}
\label{fig2}
\end{figure*}

{\bf Analytical expressions.} In order to understand which quantities determine both the shifts in the unpolarized spectrum and the $LD$, an analytical model is desirable.
 In the absence of inhomogeneities, the optical response of circular aggregates, resembling LH1 or LH2 rings, is determined by the $\ket{k=\pm1} = \frac{1}{\sqrt{2N}} \sum_n (-1)^n \exp \left( 2 \pi i k  \lfloor{n/2} \rfloor / N  \right) \ket{n}$ states, where $\lfloor{n/2}\rfloor$ is the largest integer smaller than $n/2$. These states carry all the transition dipole moment, namely $\vec{D}_{k=\pm1} \approx |\vec{d}| \sqrt{\frac{N}{2}} (\hat{x}\pm\hat{y})$, where $\hat{x}$ and $\hat{y}$ denote orthonormal axes to be identified with those shown in Figure \ref{fig2}(a). The inter-ring 
resonance coupling $V_{k,k'} = \bra{k,\vec{e}} \mathscr{H} \ket{k',\pvec{e}'}$  between excitons $k$ and $k'$ on rings centered at lattice sites $\vec{e}$ and $\pvec{e}'$, calculated by coupling all pigments on both rings according to Eq. \ref{eq1}, is much smaller than the energy difference between single-ring excitons, namely  $|\bra{k,\pvec{e}} \mathscr{H} \ket{k,\pvec{e}}-\bra{k',\pvec{e}'} \mathscr{H} \ket{k',\pvec{e}'}|$. Such condition results in a minor mixing between states $|k|\neq|k'|$,   results in an optical response of assemblies mainly determined by the coupling between the bright $\ket{k=\pm1,\vec{e}}$ and $\ket{k'=\pm1,\vec{e'}}$ states from neighbouring rings. In particular for this manifold, it can be shown that $V_{1,1}=V_{-1,-1}<0$, $V_{1,-1}=V_{-1,1}^*$ and the argument $\arg(V_{1,-1})=\xi$ for $\vec{e}-\pvec{e}'= \pm a\hat{x}$ becomes $\arg(V_{1,-1})=\xi+\pi$ for $\vec{e}-\pvec{e}'= \pm a\hat{y}$, where $a$ is the lattice spacing.

In what follows, we consider nearest-neighbor couplings on the rectangular lattice of Figure \ref{fig2}(a), as in Figure \ref{fig1B} regarding square and linear configurations, as it exhibits the optical signatures of extended delocalization while allowing simple analytical expressions. In particular, $V_{1,1}$ = -1.8 cm$^{-1}$ and $|V_{1,-1}|$ = 4.4 cm$^{-1}$. Qualitatively similar results are obtained for couplings beyond nearest-neighbours and triangular lattices, which we consider explicitly in the Supporting Information. Rings are centered at lattice sites $\vec{e}=a (q \hat{x} + r \hat{y})$, where $q=1,...,Q\in \mathbb{Z}$, $r=1,...,R\in \mathbb{Z}$ and $a=120$ \r{A} \cite{Bahatyrova_2004Nature}; single-ring excitons are labeled $\ket{\pm1,q,r}=\ket{\pm1,\vec{e}}$. Accordingly, the extended excitonic wavefunctions over the rectangular array read as
\begin{eqnarray}
\ket{\pm,k_q,k_r}& =& \sqrt{\frac{2 \times 2}{(Q+1)(R+1)}}\times\nonumber\\
& & \sum_{q,r} \sin \left( \frac{\pi k_q q}{Q+1} \right) \sin \left( \frac{\pi k_r r}{R+1} \right) \ket{\pm,q,r}\nonumber\\
\end{eqnarray}
where $\ket{\pm} =\frac{1}{\sqrt{2}} \left( \ket{1} \pm \ket{-1} \right)$ and $k_q=1,...,Q$ and $k_r=1,...,R$ are Fourier quantum numbers corresponding to the $\hat{x}$ and $\hat{y}$ directions, respectively. The optical response can be approximated according to equation \ref{Aspec2}, 

\begin{align}\label{eq:dip}
\begin{pmatrix}
|\vec{D}_{+,k_q,k_r} \cdot \hat{E}|^2 \\
|\vec{D}_{-,k_q,k_r} \cdot \hat{E}|^2
\end{pmatrix}
\propto \frac{S^2(k_q,Q)}{Q+1} \frac{S^2(k_r,R)}{R+1}
\begin{pmatrix}
 \cos^2 \phi \\
 \sin^2 \phi
\end{pmatrix}
\end{align}
where $\phi$ is the angle between the polarization of the field $\hat{E}$ and the axis $\hat{x}$. The absorption strength is determined by $S(k,W) = \sin\left(\frac{\pi k}{2}\right) \sin\left(\frac{\pi k W}{2(W+1)} \right) \csc\left( \frac{\pi k}{2(W+1)} \right)$, distributed over states $\ket{\pm}$ along orthogonal polarizations. 
Given that $S(k,W) \sim \frac{W}{k} \sin^2\left(\frac{\pi k}{2}\right) $ already for $W\gtrsim 3-4$ rings, the dipole strength becomes small for higher $k>1$ states, concentrating dipole moment in the $k_q,\, k_r=1$ states in the rectangular configuration. The energies of these states are shifted by
\begin{eqnarray}
E_\pm &=&2 V_{1,1} \left( \cos \left( \frac{\pi}{Q+1} \right) + \cos \left( \frac{\pi}{R+1} \right)  \right) \nonumber\\
       & &\pm 2 |V_{1,-1}| \left( \cos \left( \frac{\pi}{Q+1} \right) -\cos \left( \frac{\pi}{R+1} \right)  \right) \nonumber \\
       &= &\delta E \pm \Delta E.\label{eq:split}
\end{eqnarray}
Given that the absorption is composed of these two shifted optical transitions, a general shift  $\delta E =(E_++E_-)/2\propto V_{1,1}$  of the (unpolarised) absorption maximum, with respect to the single-ring case, is expected for the coherently coupled array of rings.  Such a shift from the assemblies studied in Fig.\ref{fig1B}(a)-(b), accounts for $\delta E=2V_{1,1}$  in the square ($Q=2$, $R=2$) and $\delta E=2V_{1,1}\cos(\pi/5)$ in the linear ($Q=4$, $R=1$) assemblies. Hence, the above expression  explains why the square geometry shows a greater shift than the linear assembly, with respect to the single-ring case, in the unpolarized absorption spectrum. Last but not least, note that the magnitude of $V_{1,1}$ depends also on the specific details of the structure (further details in the Supplementary Information). Hence, spectral shifts can be used as a tool to understand specific details that have been not possible to resolve by other means.


The states $\ket{+}$ and $\ket{-}$ are orthogonally polarized; polarisation along the $\hat{x}$-axis results in absorption spectra peaking at energy $E_+$ corresponding to $\vec{D}_+\propto \hat{x}$, while polarisation along the $\hat{y}$-axis results in absorption spectra peaking at energy $E_-$ corresponding to the $\vec{D}_- \propto \hat{y}$ state. The subtraction of these two spectra recorded from perpendicular polarisations,  the {\it LD}, is finite whenever $\Delta E=(E_+-E_-)/2$ does not vanish, which occurs, according to Eq.~\ref{eq:split}, strictly due to the resonance coupling $V_{k,k'}$ which breaks the circular symmetry of the assembly. Explicitly, based on Eq.~\ref{eq:split} the effect is greatest for a linear chain ($Q=1$ or $R=1$) and a saturation for both the red shift, $\delta E$, and splitting, $\Delta E$, is expected, as $\cos \frac{\pi}{Q+1}\approx 1$ already for $Q\approx 5-6$ rings.

\begin{figure*}[t!]
\includegraphics[width=16cm]{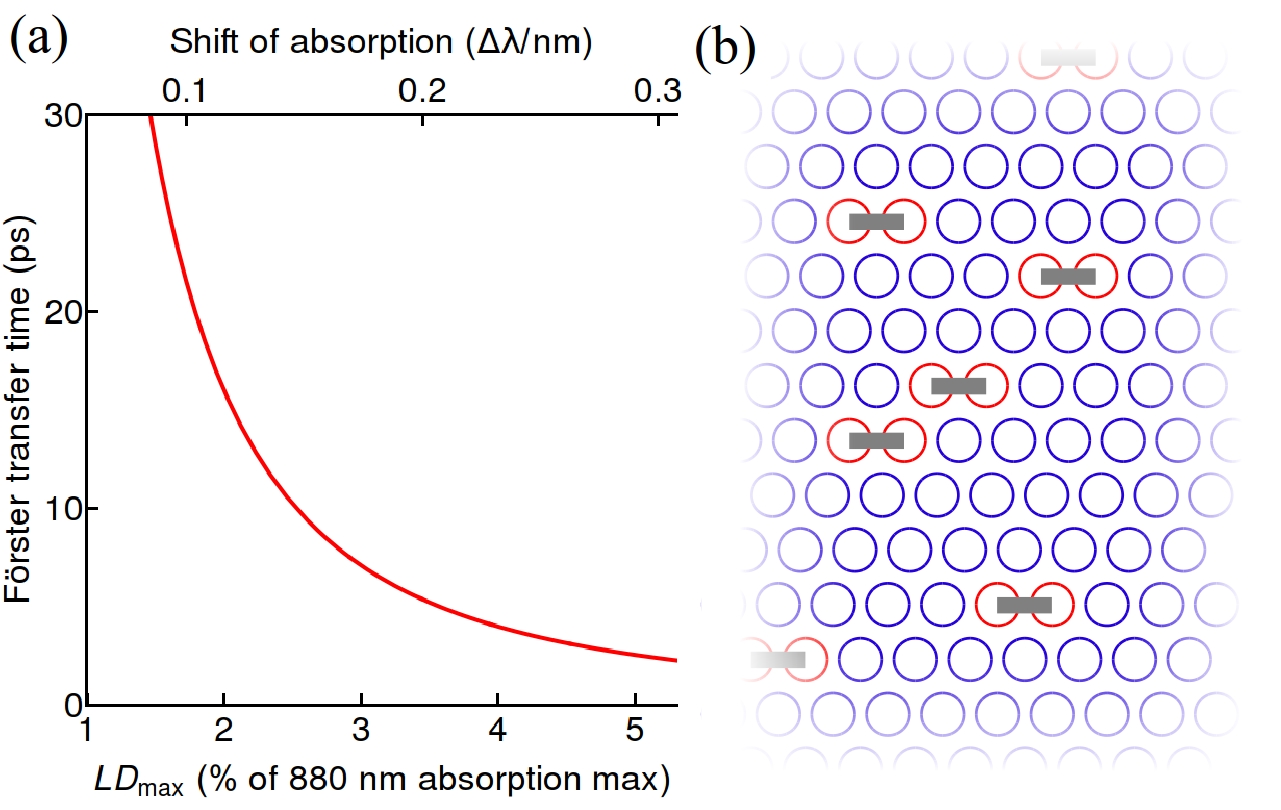}
\caption{Absorption measurements can provide direct information on the magnitude of transfer rates. In (a) is presented the transfer time between LH1 complexes as a function of the {\it LD} contrast and/or shift in the unpolarized absorption (which depend on the resonance coupling $V_{k,k'}$, itself a function of the dielectric constant). An experimental ensemble measurement of the $LD$ signal mitigates the effects of structural disorder and allows a direct absorption readout. Such an experiment is shown in (b), where ``blue" rings, mutated to maximally absorb at less than 880 nm\cite{Ghosh_88, Robert_97}, are grown in a membrane with ``red" rings, absorbing at 880 nm and linked in pairs. The aggregation of blue-shifted LH1s leads to their vanishing $LD$, while, if the axes of the red pairs align, their non-zero $LD$ signal can be measured.}
\label{fig3}
\end{figure*} 

{\bf Experimental determination of F\"orster rates.} Besides controlling the linear optical response, the resonance coupling dictates the rate at which, on a longer time scale, excitations migrate incoherently between complexes \cite{forster1965,Pullerits96}. 
The transfer among different bands which can be resolved spectrally permits experimental schemes that make use of the different spectral components in order to separate dynamical contributions \cite{Engel13a, Engel13b, Pullerits09, Hess_1995PNAS, Hildner2013}. In the case of isoenergetic transfer steps, like LH1$\rightarrow$LH1, this is not possible and it becomes hard to identify the nature
of individual dynamical components. Excitation annihilation was used to determine the transfer rate between LH2 rings in \emph{R. sphaeroides}\cite{Schubert04}, and techniques based on transient depolarization, such as anisotropy absorption recovery, have been
valuable to recognize multiple dynamical contributions at low temperatures in the core complex of {\it R. rubrum} \cite{Hunter90, Sundstrom86, Bergstrom88, Pullerits94, Fleming95, Cogdell88, sundstrom95}, but have failed to unequivocally identify components arising from intra-ring relaxation or from inter-ring energy transfer. The accurate determination of these rates is crucial for understanding the efficiency of photosynthesis.

We show here that the possibility to obtain the resonance coupling between isoenergetic species of LH1 or LH2 complexes-- through the absorption measurements illustrated in Figure \ref{fig2}-- circumvents the ambiguity in these isoenergetic landscapes and opens up a promising experimental scheme to quantify their mutual transfer rates. 

Two facts of major importance are that the shift of the unpolarised  absorption is $\delta E=V_{1,1}\cos (\pi/(Q+1))$ and that the splitting between bright tranitions is $2\Delta E=4 |V_{1,-1}|\cos (\pi/(Q+1))$ for linear geometries. Accordingly, measurement of the shift in the unpolarized absorption spectrum determines $V_{1,1}$ while, as will be shown shortly, the determination of $V_{1,-1}$ through an $LD$ measurement is possible.
A good quantitative agreement between the full calculated spectrum and the subtraction of two Lorentzian lineshapes with full width $\Gamma \approx$ 465 cm$^{-1}$ (in agreement with experiment) separated by $V_{1,-1}$ permits to extract the relation
\begin{equation}
LD_{max}= C_1 \, \frac{|V_{1,-1}|}{\Gamma}+O\left( \left(\frac{|V_{1,-1}|}{\Gamma}\right)^3 \right)
\end{equation}
for the $LD$ amplitude, $LD_{max}$, with a constant $C_1=\frac{3\sqrt{3}}{2}$ in the case of two rings, which accounts for the geometrical details of the array.  Accordingly, it is possible to determine through the unpolarized spectrum and the $LD$ amplitude, the values of $V_{1,1}$ and $V_{1,-1}$ independently.

Generalized F\"orster theory is often used to calculate the rate of incoherent transfer $\gamma_{k \rightarrow k'}$ from a donor exciton $k$ to acceptor exciton $k'$
\begin{align}
\gamma_{k \rightarrow k'} &= \frac{2\pi}{\hbar} |V_{k,k'}|^2 \, Z_k \, I_{k,k'},
\end{align}
where $V_{k,k'}$ is, as before, the coherent excitonic coupling among single-ring states, $Z_k$ is the thermal population (Boltzmann factor) and $I_{k,k'} = \int_0^\infty F_k(\epsilon) A_{k'}(\epsilon) d\epsilon$ denotes the spectral overlap integral of the donor fluorescence from exciton $k$ and acceptor absorption of exciton $k'$, $F$ and $A$, respectively\cite{Silbey04,Scholes03}. The total transfer rate from donor to acceptor is $\gamma= \sum_{k,k'} \gamma_{k\rightarrow k'}$. Although we have taken a large inhomogeneous broadening in our calculations, there is experimental evidence of a predominance of homogeneous broadening of LH1 at room temperature\cite{Scholes03, Freiberg92, Kramer84, Rijgersberg80}. Under the assumption that the emission is also spectrally homogeneous, the total rate is $\gamma \approx \sum_{k,k'=\pm1} \gamma_{k \rightarrow k'}$, and the F\"orster rate can be calculated as
\begin{eqnarray}
\gamma &=& \frac{2\pi}{\hbar} \left( 2|V_{1,1}|^2 + 2|V_{1,-1}|^2 \right) Z_1 I \label{eq:FRex}
\end{eqnarray}
where $I=\int_0^\infty F(\epsilon) A(\epsilon) d\epsilon$ is the full spectral overlap and $Z_1$ is the thermal population of the $\ket{k=\pm1}$ states, which depends on the single-ring model.

As shown above, $V_{1,1}$ = -1.8 cm$^{-1}$ and $|V_{1,-1}|$ = 4.4 cm$^{-1}$ are quantified by shifts in absorption and non-zero {\it LD} contrast, respectively, which allow a calculation of the F\"orster rate with a minimum of model input. The dipole moments $|\vec{d}|$ and the dielectric contant $\kappa$ = 2 predict an inter-ring transfer time of 8 ps, which is well within the bounds ($\sim$1ps -- 20ps) of estimates from current experiments and theoretical calculations \cite{Sundstrom86,Ritz01,sundstrom95}. Determination of the inter-ring coupling strength by measurement of the {\it LD} thus provides a new method to determine F\"orster transfer rates. This orocedure would then provide further insight into the values of the dielectric constant and the BChl dipole moment {\it in vivo}, which are parameters usually estimated indirectly. Figure \ref{fig3} (a) plots the transfer time, {\it LD} contrast and absorption shift as a function of these quantities according to Eq. \ref{eq:FRex}.

Several elements have to be brought together in an experiment which confirms excitonic delocalization across extended domains. A speculative setup is illustrated in Figure \ref{fig3} (b) based on mutagenesis of blue-shifted LH1s \cite{Ghosh_88, Robert_97}. If a link between LH1s by means of mechanical bridges or affinity domains is accomplished, the formation of linear-like assemblies will be possible. Additionally, these assemblies must be \emph{macroscopically} aligned in order to set the polarisation excitation directions. The blue-shifted LH1s will serve then to build the membrane scaffold to accomodate the linear arrays of non-mutated LH1s and fix their orientations. Fluorescence yield detection upon polarised excitation performed on single assemblies \cite{Ghosh_Biochem_2003} is also a promising technique to confirm/disregard the hypothesis of long-range excitonic delocalisation in natural LH aggregates.

\section{Conclusions}
In summary, we have modeled and characterized, through analytical expressions and numerical calculations, the optical response due to resonance coupling between multiple rings in photosynthetic membranes of purple bacteria. This investigation highlights the importance of a description, at room temperature, of light absorption beyond the standard assumption of excitons restricted to individual rings. The dipole moment redistribution that emerges in 2D arrays of LH1 rings, due to extended excitons, leads us to propose an experimental procedure, based on polarized and unpolarized absorption spectra of small linear-like arrays of rings, which quantifies inter-ring resonance coupling. We show that this experimental procedure allows an alternative, indirect measurement of their incoherent F\"orster transfer rate, which carries additional information on currently poorly-characterized parameters, like the dielectric constant or the {\it in vivo} BChls dipole moment. Such a measurement could be accomplished through absorption of an ensemble of oriented linear-like assemblies or through fluorescence yield detection of single assemblies which have been excited by polarized fields.

{\bf Acknowledgement}\\
This work was funded by the EU STREP projects PAPETS and QUCHIP, the ERC Synergy grant BioQ and an Alexander von Humboldt Professorship. Additional support was provided by the National Science Foundation through PFC@JQI and the National Science Foundation Graduate Research Fellowship Program under DGE-1322106. Any opinions, findings, and conclusions and recommendations expressed in this material are those of the authors and do not necessarily reflect the views of the National Science Foundation. The authors would like to thank R. Ghosh (University of Stuttgart), F. Jelezko (University of Ulm) and Shai Machnes (Weizmann Institute, Israel) for discussion at early stages of this work, and R. Ghosh for experimental data and a careful and critical reading of the manuscript.

Supporting information includes a study of the effects of inhomogeneous broadening on measures of exciton delocalization and a generalization of the results to alternate membrane geometries.

\bibliography{biblio}

\clearpage

\appendix


\section{Hamiltonian of LH1 complex}\label{Hcomplex}
\begin{figure}[htbp]
\includegraphics[width=1\columnwidth]{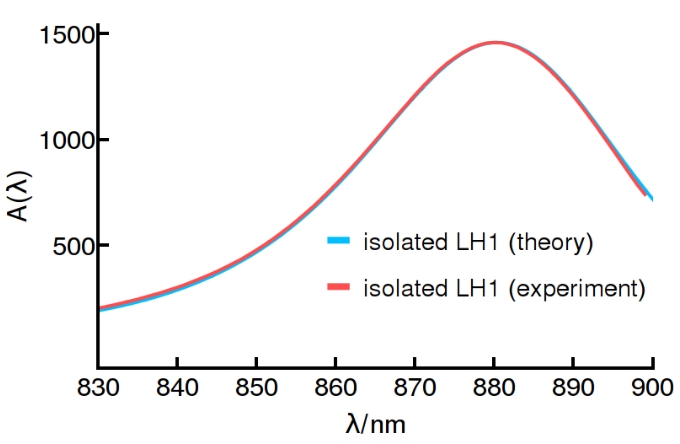}
\caption{The Hamiltonian and noise parameters taken in this study reproduce the experimentally-observed absorption spectrum.}
\label{figLH1}
\end{figure}
In this study we have employed a model of the {\em R. rubrum} LH1 where 32 BChl molecules, bound to 16 $\alpha$ and $\beta$ polypeptides as $\alpha\beta$(BChl)$_2$ subunits, are equidistantly arranged along a ring of radius $r$ = 47 \r{A}. Dipole moments are taken to be tangential to, and in the plane of, the ring. The sub-nanometer distance between neighbouring chromophores implies that the nearest-neighbour couplings depend on the geometry of the electronic wave-function of each chromophore and cannot be inferred directly from a dipolar interaction in the point-dipole approximation. Couplings on the ring can be fitted using fluorescence anisotropy measurements, resulting in nearest-neighbor couplings $Q_1= 600$ cm$^{-1}$ and $Q_2 = 377$ cm$^{-1}$ for the intra- and inter-dimer couplings, respectively \cite{Timpmann_ChemPhysLett2005}. As discussed in the text, homogeneous broadening is taken into account using either a Lindbladian master equation with a pure dephasing rate of $\gamma_d$ = 106 cm$^{-1}$ or the equivalent protocol of dressing excitonic stick-spectra with a Lorentzian of full-width half-maximum $2\times\gamma_d$. These parameters reproduce the experimentally-measured absorption spectrum of isolated LH1, see Fig.\ref{figLH1}. Coupling between dipoles on different rings, separated by $a$ = 120 \r{A}, is calculated using the dipole-dipole interaction
\begin{align}
    J_{n,m}=\frac{1}{4\pi\kappa} \left(
    \frac{\vec{d}_{m}\cdot\vec{d}_{n}}{\left| \Delta \vec{r}_{n, m} \right|^3}-
    \frac{3\left( \vec{d}_{n}   \cdot \Delta \vec{r}_{n,m} \right)
           \left( \vec{d}_{m} \cdot \Delta \vec{r}_{n, m} \right)}
     {\left| \Delta \vec{r}_{n, m} \right|^5}\right), \label{eq:dipole_dipole}
\end{align}
where $\Delta \vec{r}_{n, m}=\vec{r}_{n}-\vec{r}_{m}$ and $\kappa$=2 is the relative permittivity. The LH1 dipoles are taken to be of magnitude $\sqrt{2.4}$ $\times$ 6.3 D = 9.8 D to take into account the observed hyperchromism \cite{Ghosh_88}.

\section{Measures of extended delocalization}
Excitonic properties from purple bacteria membranes can be obtained from the Coulomb exchange Hamiltonian
\begin{align}
H&=\sum_m^{2NR}(\epsilon_m+\delta_{\epsilon_m})\ket{m}\bra{m}\nonumber\\
& +\sum_{m\ne n}^{2NR}J_{mn}(\ket{n}\bra{m}+\ket{m}\bra{n})\label{HexcS}
\end{align}
where $J_{nm}$ denotes the interaction strength between the induced transition-dipoles of pigments $n$ and $m$, corresponding to electronic excited states $\ket{n}$ and $\ket{m}$ with energies $\epsilon_n$ and $\epsilon_m$. Note that each ring contains $2N=32$ pigments and we examine a line of rings ($Q$=1), as in the main text. The wavefuntions take a remarkable simple form in the absence of protein inhomogeneities ($\delta_{\epsilon_m}=0$), as given in eq (5) of the main text. 

Explicitly, the bright states of relevance for the optical response are 
\begin{align} 
\ket{\pm,k_x=k_y=1,Q=1} \propto \sum_r \sin \left( \frac{\pi}{R+1} \right) \ket{\pm,r}. \label{eq:wave}
\end{align}
These states present populations in each ring proportional to $\sin^2( \pi r/(R+1))$ which is largest for the rings lying in the middle of the linear chain.   Note that in the abscence of coherent interaction among rings, all states should have populations concentrated in single rings. As a consequence, on average, a density operator will present a statistical mixture with populations being equally shared among all rings. 

\begin{figure*}[ht!]
\centering
\includegraphics[width=16cm]{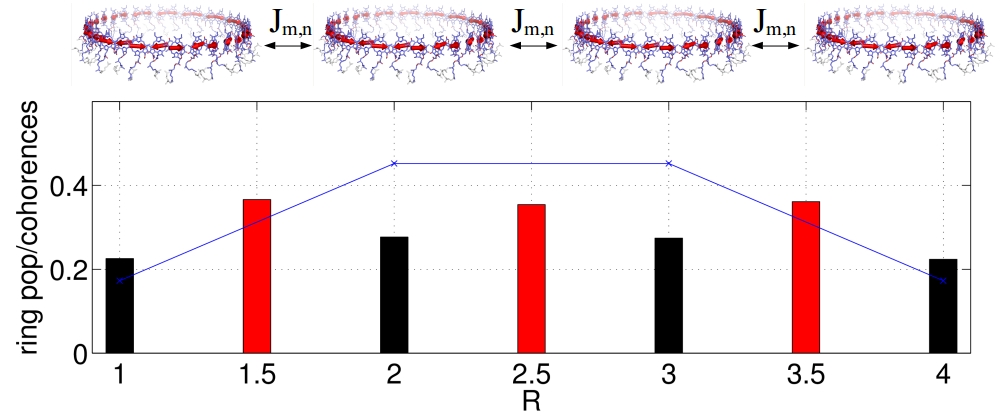}
\caption{The linear array of rings in top presents ring populations $\sum_{n\epsilon r}|\bra{n}\rho\ket{n }|$ and total inter-ring coherences $\sum_{n\epsilon r_1,m\epsilon r_2}|\bra{n}\rho\ket{m }|$ shown in black and red bars, as a result from averages of 50000 stochastic realisations of inhomogeneous noise. The result of the the noise-less model for populations is presented in blue connected crosses.}  
\label{fig1A}
\end{figure*}

\begin{figure}[htbp]
\includegraphics[width=1 \columnwidth]{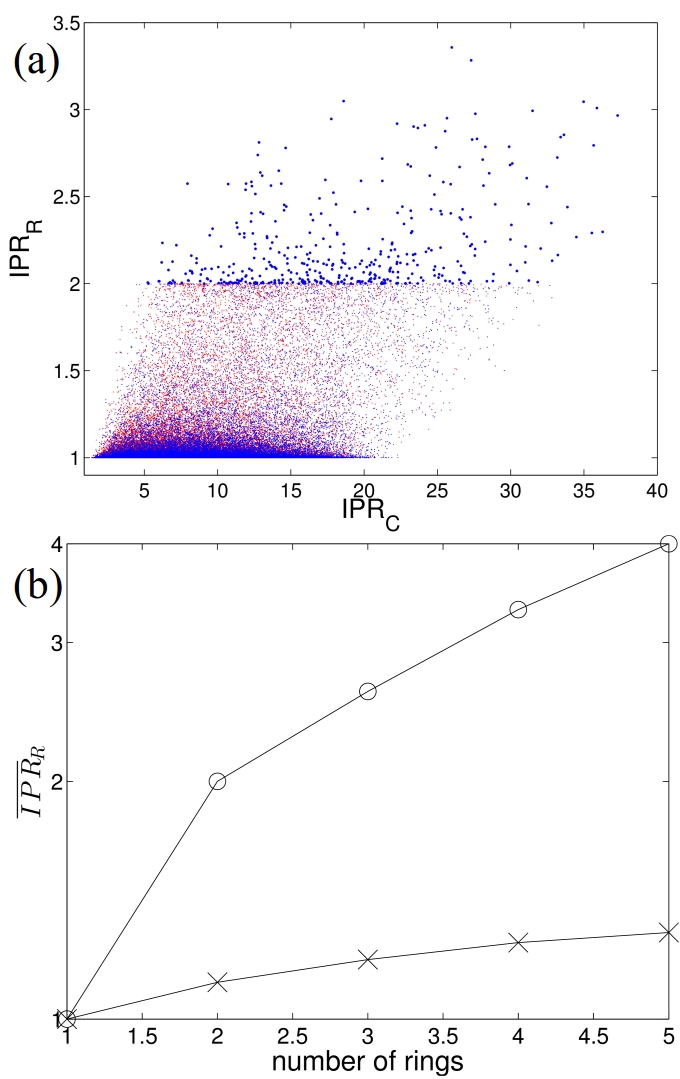}
\caption{(a) Scatter plot (points) of the $IPR_C$ and $IPR_R$ from eqs.(\ref{iprC},\ref{iprR}) for two (red) and five (blue) rings, for states with the greatest dipole moment in realisations of the quasi-static fluctuations in eq \ref{HexcS}. States that delocalize beyond 2 rings are highlighted. In (b) is presented the average $\overline{IPR}_R$ for the brightest state in each noise realisation (crosses) and for the noiseless wavefunction in eq (\ref{eq:wave}) (circles), as a function of the length of the array. Symbols connected by straight lines to guide the eye.}
\label{fig2A}
\end{figure}

Absorption is a linear functional in the density operator. Hence, we concentrate on the density operator $\rho=\overline{\ket{\alpha}\bra{\alpha}}$ which results from averaging the eigenstates of eq (\ref{HexcS}) under realistic conditions, i.e. with appropriate magnitude for protein-induced fluctuations $\sqrt{\langle \delta_\epsilon^2 \rangle}$ = 325 cm$^{-1}$. Our choice of magnitude of these fluctuations is justified by a detailed analysis of absorption spectra of LH1 in {\it R. rubrum}, see Fig. \ref{figLH1}. 

The average population of each ring for the subset of states $\ket{\alpha}$ with the largest light-induced dipole strength, is presented in Figure \ref{fig1A}. Here, a non-uniform population in the linear configuration of rings can be seen, where the complexes in the middle of the chain present a greater population compared to the rings at the chain edges. The pattern of ring populations displayed in the presence of inhomogeneities resembles the noiseless case from states \ket{\pm,k_x=k_y=1,Q=1}, which illustrates the formation of excitons fulfilling the full array boundary conditions in the presence of quasi-static fluctuations. The resilience of the excitons to the inhomogeneous disorder illustrates that the disorder produce noticable smearing out of excitonic delocalization, but does not completely erase its features. Another indicator of excitonic coherence among neighbouring rings is also presented in Figure \ref{fig1A}, namely the total coherence  $\rho_{r_1,r_2}=\sum_{n\epsilon r_1,m\epsilon r_2}|\bra{n}\rho\ket{m }|$ among all chromophores that belong to {\it neighbouring}  rings $r_1$ and $r_2$. This total inter-ring coherence, ussually assumed to be negligible for inter-ring exciton transfer, is manifestly non-zero in Figure \ref{fig1A}, and is the cause for the absorption anisotropy that produces the finite linear dichroism ({\it LD}) discussed in the main text. 

To characterise the excitons formed upon photon absorption, useful quantities are the conventional inverse participation ratio 
\begin{align}
IPR_C\equiv \left(\sum_i^{2NR} |c_{i}^\alpha|^4\right)^{-1} \label{iprC}
\end{align} 
and a generalisation which we denote the ring inverse participation ratio 
\begin{align}
IPR_R\equiv \left(\sum_{r=1}^R \left(\sum_{i\epsilon r}^N |c_{i}^\alpha|^2\right)^2\right)^{-1} \label{iprR}
\end{align}
where the inner and outer sums are performed over BChls that belong to a specific ring $r$ and over all rings in the array, respectively.

As defined, $IPR_C$ ranges from 1 to the total number of pigments in the array and measures how many chromophores participate in a given exciton, while $IPR_R$ ranges from 1 for any exciton confined to a single ring, to $IPR_R=R$ for pure states that are evenly delocalised over the entire array of $R$ rings. For $IPR_R> 1$ or $IPR_C> 32$, these two quantitites demonstrate unambiguously the presence of bright excitons over domains greater than a single ring (see Figure \ref{fig2A}(a)). A  general trend shows larger exciton lengths for arrays with higher number of rings.  Figure \ref{fig2A}(a)  highlights realisations extending over more than 2 rings ($IPR_R>2$), which extend well beyond the current paradigm of single ring excitons (lying flat with $IPR_R=1$),  sufficient for the description of processes with a longer built-in timescale.

The dependence of excitons length with the size of the array is made conspicuous in Figure \ref{fig2A}(b) where the average $\overline{IPR}_R$ is shown as a function of the linear ($Q=1$) array size for the two brightest states $\ket{\alpha}$ of each realisation. Here, it can be appreciated an increase in exciton size with the array extension in the same qualitative manner as the homogeneous wavefunctions of eq (\ref{eq:wave}) (also shown in the same figure). 

Based on the observations that the wavefunction and the excitonic length from averages of inhomogeneous noise resemble qualitatively the features that arise from a noiseless treatment, it can be safely assumed that the inhomogeneous broadening is included, the noiseless features are not completely smeared out. It becomes natural therefore, to assert the properties of the system based on the noiseless characterization presented through equations (5)-(7) in the main text.

\section{Shift of the unpolarised absorption spectrum: dependence of $V_{1,1}$ on the geometry of the ring dipoles}
\begin{figure}[t!]
\centering
\includegraphics[width=7cm]{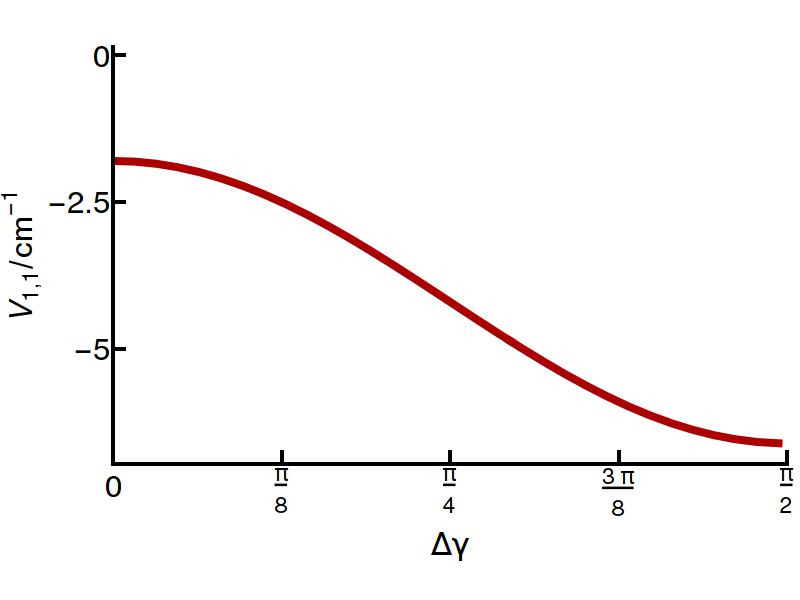}
\caption{Coupling $V_{1,1}$ as a function of the angle $\Delta\gamma$ between the $Q_y$ transition dipole and the ring tangent.}
\label{fig:v11}
\end{figure}

The resonant interaction betwee rings in general assemblies can result in a shift of the absorption spectra, addressed by equation 6 in the main text and numerically supported by Figure 2(b) and 2(d) in the main text. The magnitude of this shift, in comparison to the isolated ring situation, depends on the structural details of the rings. As it has been stated in section \ref{Hcomplex} of this SI, we have used a structure where the $Q_y$ transition dipoles are tangent to the circumference of the ring. However if this angle, $\Delta\gamma$, is increased, the magnitude of $V_{1,1}$ increases, as presented in Figure \ref{fig:v11}. Accordingly, the determination of the shift in the spectra reflects specific details of the rings, which escape measures performed on single rings, but are otherwise accessible through the observation of assemblies.

\section{ Inhomogeneous broadening.}
\begin{figure*}[t!]
\centering
\includegraphics[width=16cm]{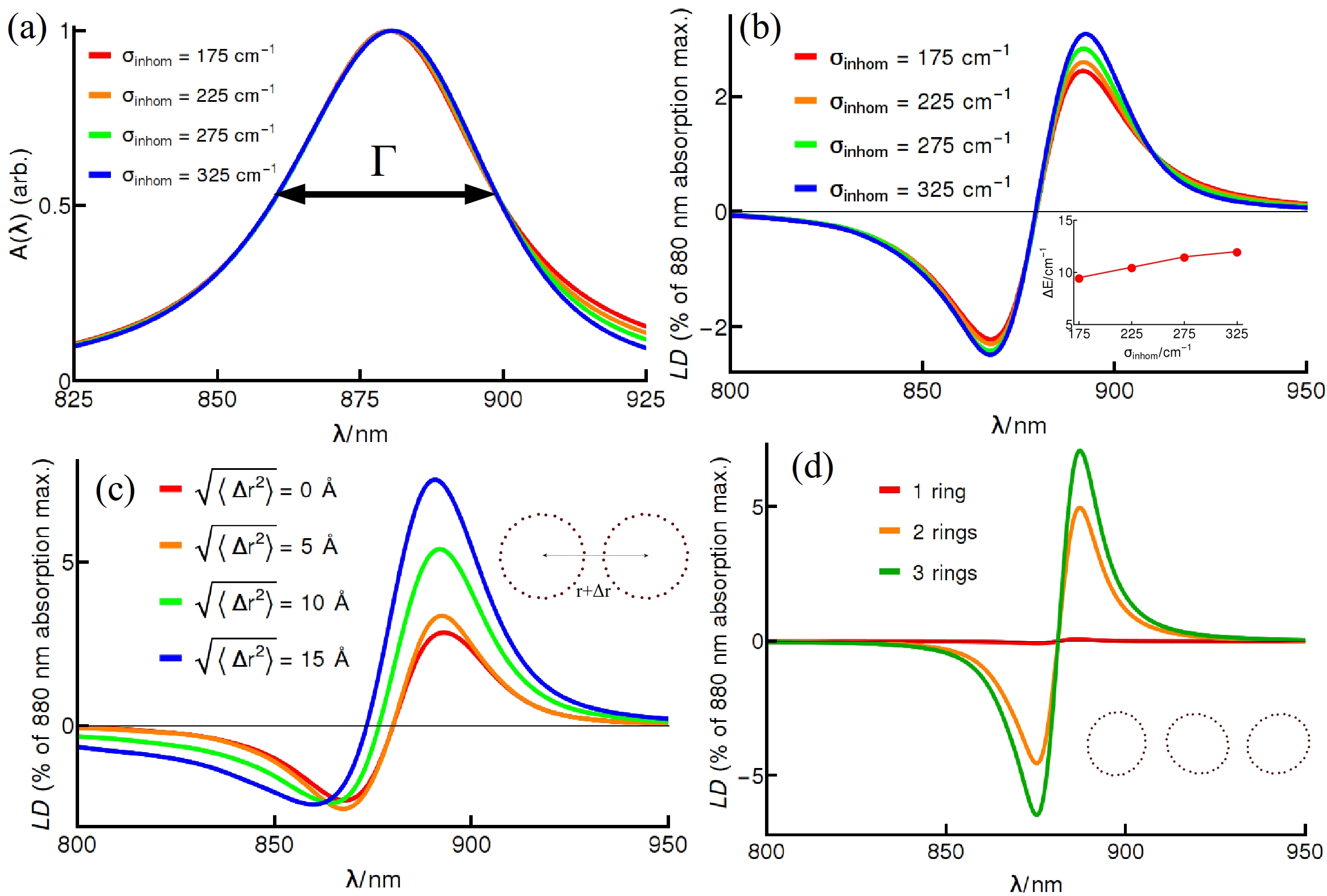}
\caption{Effects of energetic and structural disorder on $LD$. In (a), the overall absorption spectra takes on more Gaussian lineshape,  manifested from steeper tails of the spectra as inhomogeneous disorder is increased.  The splitting between orthogonally-polarized manifolds, $\Delta E$, in (inset b), has only a slight increase with inhomogeneous noise, which results in an $LD$ signal (b) robust to energetic disorder ($\sigma_{inhom}$ = 175 - 325 cm$^{-1}$). The $LD$ is also robust to variations of the inter-ring distance, as shown in (c). In (d) the $LD$ ensemble averaged over randomly-oriented oval structures in linear arrays with 1, 2 and 3 rings.}
\label{fig:pert}
\end{figure*}

The spectral lineshape and width are determined by both homogeneous and inhomogeneous contributions. Two effects are manifested when inhomogeneities are introduced in excitonic transitions, namely, lineshape modifications and energy shifts. Figure \ref{fig:pert}~(a) shows that overlapping spectra are obtained with different combinations of homogeneous and inhomogneous broadening, being the most conspicuous signature of their slightly different lineshapes, the difference in the low energy tails. For the case of smaller inhomogeneous broadening, a Lorentzian-like lineshape appears  with slower rise as compared to the result with higher inhomogeneity with a Gaussian-like lineshape. For values of inhomogeneous disorder bounded by the full spectral width, the energy splitting between absorption spectra along orthogonal polarizations is greater with the inhomogeneous noise Figure \ref{fig:pert}~(b). This phenomenon can be understood by the excitonic repulsion which results in slightly larger splittings for greater levels of noise \cite{Silbey_JCP2001}. Accordingly, the $LD$ changes due to the inhomogeneous noise (see inset of Figure \ref{fig:pert}~(b)), with a very counter-intuitive result, namely, greater inhomogenities produce greater $LD$ amplitude. Such increase can only be traced back with full numerical simulations, as it requires to account for all the dark levels in order to quantitatively assert the shift calculated in the inset of Figure \ref{fig:pert}~(b). Note however that within the inhomogenities expected in these systems, the energy difference among orthogonal absorption spectra does not produce dramatic changes in the $LD$, given that we have been careful to reproduce in all cases the absorption profile  shown in Figure \ref{fig:pert} (a).   The major change in the $LD$ will be due to the full broadening  of the optical transition transition --labeled in Figure \ref{fig:pert}~(a) as $\Gamma$-- as it becomes conspicuous by comparison of the $LD$ from homogeneously broadened spectra with  a width of 2$\times$106 cm$^{-1}$  in Figure 1 of the main text, and the $LD$ of full spectrum, here presented. In the former case, the amplitude of the $LD$ can reach about $5\%$ of the absorption spectrum, while in the latter case the $LD$ amplitude is reduced quite generally to $2-3\%$, as Figure \ref{fig:pert}~(b) illustrates. It can be  concluded from this analysis, that the LD changes only slightly depending on the inhomogeneous component in the spectra with a robustness only vulnerable to the total width $\Gamma$ of the absorption spectrum.  

The noise model presented above accounts for environmentally-induced inhomogeneities in the pigment energies. However, it does not account for explicit  perturbations of the chromatophore vesicle, such as variations in the inter-complex distance or for structural perturbations, such as elliptical deformations of the rings.

Examination of paracrystalline \cite{Scheuring_2005Science} domains of LH2 in \emph{Rsp. photometricum} and close-packed orthorhombic crystals \cite{Ghosh_1990,jamieson2002} in \emph{R. rubrum} imply variations of only a few Angstroms in the center-to-center distances among neighbouring ring structures \cite{scheuring2004,Scheuring_2005Science}. 
Figure \ref{fig:pert}(c) shows the effect on the {\it LD} when the inter-complex distance suffers stochastic variations which are much greater than the variations observed from pair correlation functions calculated from atomic force microscopy (AFM) experiments ($\sqrt{\langle\delta r^2\rangle}=0,5,10,15$ \r{A} as compared to $\sqrt{\langle\delta r^2\rangle} \sim 1$ \r{A}  observed in \cite{Scheuring_2005Science}). This figure illustrates that the contrast in the LD actually increases with greater center-to-center distance variations. This surprising result can be understood from the functional form of the Coulomb interaction. At a center-to-center distance of a few tens of nanomenters, the Coulomb interaction scales as $1/r^3$ which means that bringing rings together by $\delta r$ increases the coupling more than it is decreased by separating them by $\delta r$, hence, the $LD$ is enhanced in average.  A further signature is apparent for large variations reflected by the red shift of the $LD$. This signature can be traced back to the amplitude of absoprtion along $\hat{x}$ and $\hat{y}$ axes, which for strong variation of the center-to-center distance makes the former absorption greater than the latter, and hence produces a $LD$ which is slightly asymmetric.  Note, however, that the few Angstrom variations observed in experiments are too small to appreciably affect the $LD$ signal by its magnitude or to enhance appreciably the absorption along one of the polarization direction with respect to the orthogonal, by means of inter-ring distance inhomogeneities.

In general, LH complexes in purple bacteria express geometries which may differ from the circular ring of \emph{R. rubrum} in our model. Another conformation of the LH1 in \emph{R. rubrum} has been observed to be significantly ellipsoidal \cite{jamieson2002}. In addition, the LH2 of \emph{Rubrivax gelatinosus}, obtained by AFM \cite{scheuring2001}, and the latest LH1 X-ray structure of \emph{Tch. tepidum} both exhibit a degree of ellipticity\cite{Miki_2014Nature}. It is not known if the long axes of elliptical rings may be aligned or randomly oriented along the membranes where they are synthesized. A study of elliptical structures is therefore in order as it might clarify the scope of the $LD$ to witness extended delocalization.

 Oval structures with an apparently small eccentricity $\sqrt{1-\left( \frac{a}{b} \right)^2}=0.32$, where $a$ and $b$ are the minor and major axes, respectively, produce energy shifts among $k=\pm1$ states of $\simeq 100$ cm$^{-1}$ in a single ring \cite{Gerken_JPC2003}, which is 10 times larger than the shifts due to inter-ring excitonic coupling, presented in the inset of Fig.~5~(b). In these experiments the excitation polarisation was fixed with respect to the major axis of the single elliptical ring. In an ensemble, there is no reason to believe that the orientations of major axes will or will not correlate. In order to isolate the effect from elliptical deformation, we momentarily set the pigment energy inhomogeneities to zero.
Now, if an ensemble presents randomly-oriented major axes and the rings do not couple coherently, then the finite LD signals, stemming from the elliptical geometries, cancel out and lead to a vanishing LD (see Fig. \ref{fig:pert} (d)). In order to account for the elliptical deformations we use an interaction between nearest neighbours with a functional form that only depends on their relative distance $J_{n,m}=\frac{\Delta r_0^3}{\Delta r_{n,m}^3} Q_{1,2}$ where $r_0\approx 9.2\AA$ is the distance among pigments in the circular geometry, and $Q_{1,2}$ are the nearest-neighbour couplings. In the same fashion, if an ensemble of randomly-oriented coupled rings is considered, then the LD arising from the elliptical perturbation vanishes and only the nonzero signal arising from the coherent interaction remains. Figure 4(e) shows that the nonzero LD for coherently-coupled linear arrays is robust to the ensemble-averaged elliptical perturbation, as it results in a maximum value of 5 \% for two rings, which, as expected, is similar to the value, with just homogeneous noise, presented in Fig. 1 of the main text.

It was observed that the macromolecular assembly in purple bacteria vesicles can indeed present long-range organisation \cite{vanGrondelle_PNAS2004}. Therefore, the likelihood of having affinity domains in each ring that set specific connecting residues to neighbouring units and result in an overall specific orientation of all major axes, is not out of the question. If the macromolecular assembly has a preferred orientation of rings-ellipse axes, a finite $LD$ will be measured even in the absence of appreciable inter-ring excitonic delocalization.

\begin{figure}[htbp]
\includegraphics[width=1\columnwidth]{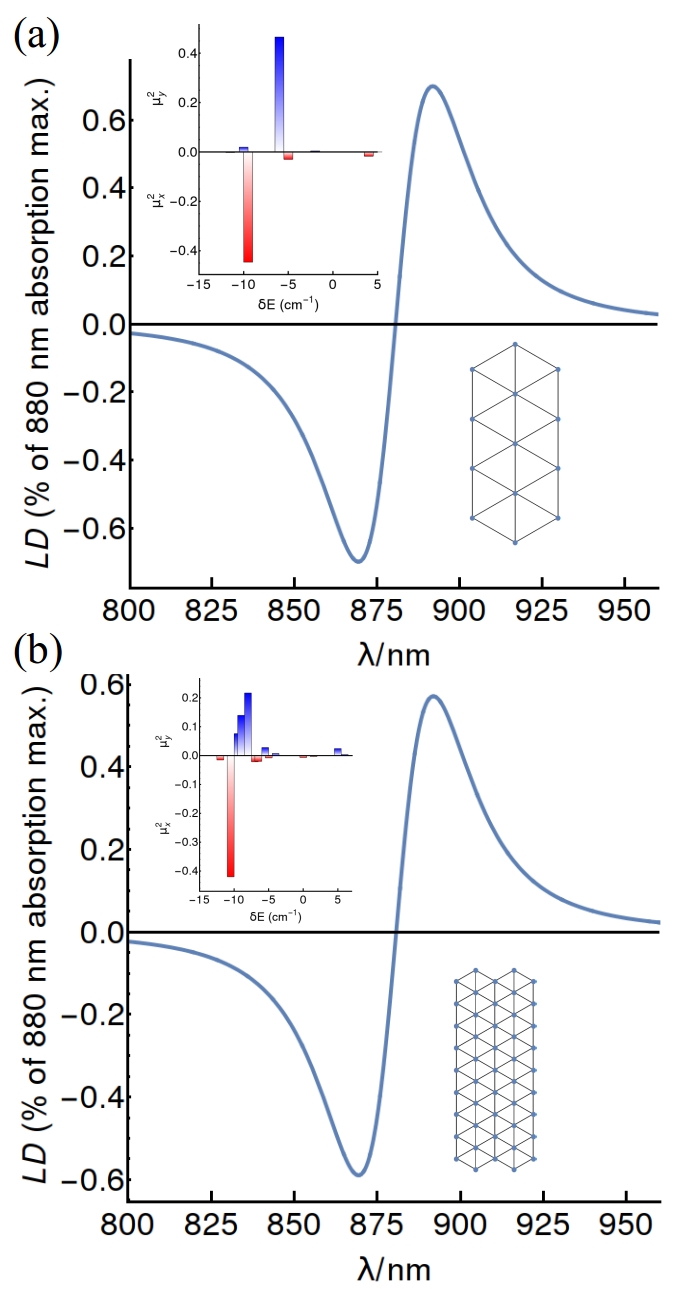}
\caption{{\it LD} spectra for the triangular lattice configuration and increasing array size. Insets show the lattice configuration studied and stick polarized specta. A small asymmetric array, as in (a), exhibits a larger {\it LD} than its double (b) due to a decreased splitting of polarized states. This can be qualitatively understood from the results for the square unit-cell, eq (7) of the main text.}
\label{fig3A}
\end{figure} 

\begin{figure}[htbp]
\includegraphics[width=1\columnwidth]{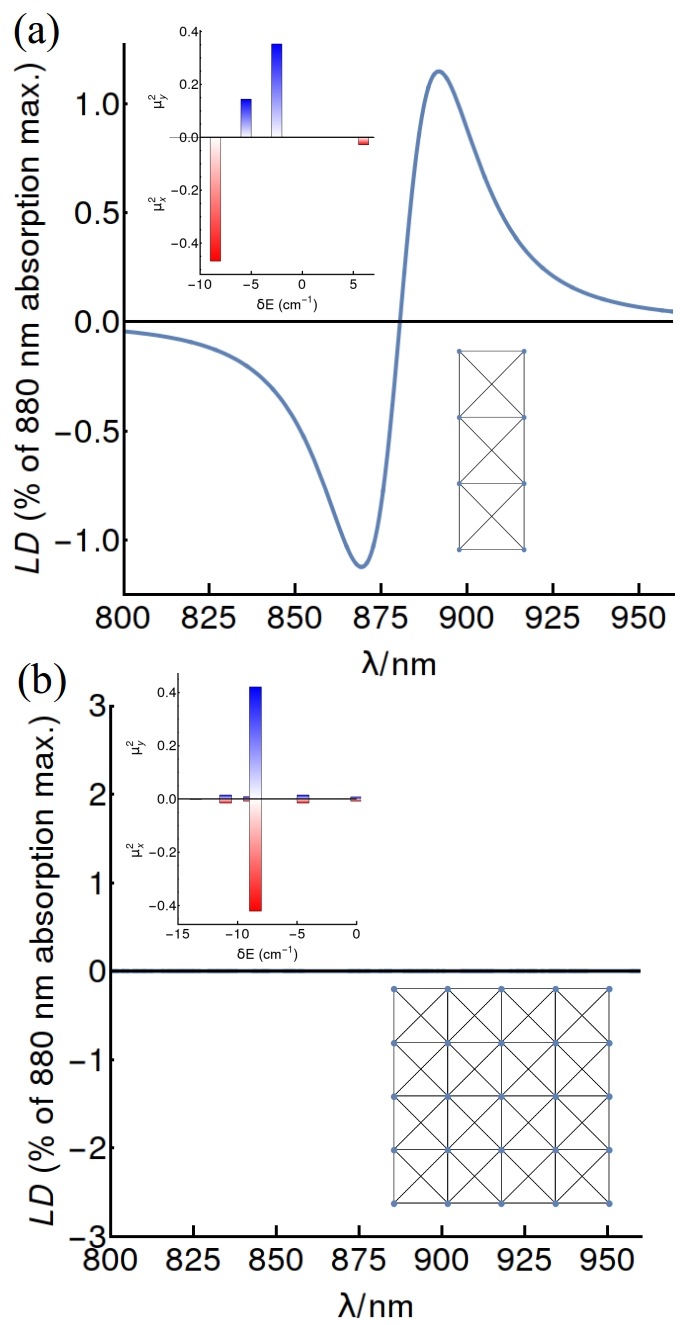}
\caption{{\it LD} spectra for the square lattice configuration and non-nearest-neighbor interactions. Insets show the lattice configuration studied and stick polarized specta. The strip of dimers in (a) reflects the native organization of LH1s in membranes containing LH1 and LH2, and exhibits a non-zero {\it LD}. Symmetric arrays ($Q=R$), as in (b), show zero {\it LD} since the polarized states are degenerate in energy. These results are qualitatively similar to those for the square unit-cell with nearest-neighbor interactions, eq (7) of the main text.}
\label{fig4A}
\end{figure}

In general, if  the geometrical contribution to the $LD$ is independent of the size of the assembly, then a simple subtraction among $LD$ signals, $\Delta LD$,  arising from different assemblies lengths will therefore represent the residual contribution resulting from the extended excitonic delocalisation among such assemblies.

\section{General 2D geometries}\label{several symm and large}

The description in the main text of excitonic delocalisation over extended domains made use of square unit-cell lattices. It was concluded that arrays with a large aspect ratio and small width are desirable in order to obtain an polarized optical response which encodes the symmetries of the extended excitons. Here we generalise these statements to arrays with other unit cells. Specifically, we analyse a triangular lattice and, to prove the generality of our statemements, we also study a square unit cell with non-nearest-neighbour interactions. 

The possibility to use the natural assembly of these complexes to probe inter-ring excitonic delocalization is investigated by numerical diagonalization of the Hamiltonian in the manifold $|k|,|k'|=1$ for a triangular unit-cell lattice (Figure \ref{fig3A}). The triangular unit cell approximates the natural ``quasi-hexagonal" aggregation state observed for LH1 complexes in photosynthetic  membranes of {\it R. rubrum}\cite{karrasch95}. Triangular para-crystaline domains of LH2 complexes also form in {\it Rsp. photometricum} at low light intensity conditions \cite{scheuring2006}. Figure \ref{fig3A}(a), shows that the dipole strength in the triangular lattice is concentrated at the band edges, analogous to the prediction made by eqs (3-5) in the main text. As shown in this figure, states polarized along and perpendicular to the long-axis of the array are split in energy, leading to a finite {\it LD} once the stick spectra is dressed with appropriate line-shape functions. Figure \ref{fig3A}(b) highlights another important feature. Here, even though an asymmetric array is being considered with the same aspect ratio, the {\it LD} signal is, in fact, smaller ({\it LD}$_{max}$ $\approx$ 0.6 $\%$ in (b) as opposed to {\it LD}$_{max}$ $\approx$ 0.7 $\%$ in (a)). This can be understood from eq (7) in the main text, which shows that the energy splitting between bright states decreases for arrays of the same aspect ratio but increasing size. Hence, independent of unit-cell configurations, only arrays with a width of a few rings will be suitable to probe extended excitonic delocalization through {\it LD}.

The same conclusion can be drawn based on a square unit-cell with non-diagonal interactions, as shown in Figure \ref{fig4A}. The additional interactions produce additional splittings, and arrays with appreciable aspect ratio ($Q/R\gg 1$ or $Q/R\ll 1$) still present a dipole moment redistribution that leads to polarization anisotropy. If the array is symmetric, the polarized states are not split in energy and the array shows zero {\it LD}, which can be seen from eq (7) in the main text. These results show that, independent of the unit-cell geometry, it is possible to witness {\it LD} in general arrays of small width and appreciable aspect ratio.



\end{document}